\documentclass[epj,final]{svjour}
\usepackage{graphics}

\begin{document}

\title{Structure determination, valence, and superexchange 
in the dimerized low temperature phase of \mbox{$\alpha$'-NaV$_2$O$_5$}}
\titlerunning {Structure, valence, and superexchange in \mbox{$\alpha$'-NaV$_2$O$_5$}}

\author{A. Bernert\inst{1} \and T. Chatterji\inst{1,2}
\and P. Thalmeier\inst{3} \and P. Fulde\inst{1}}
\institute {
Max-Planck-Institut f\"ur Physik komplexer Systeme, 01187 Dresden, Germany
\and Institut Laue-Langevin, BP 156, F-38042 Grenoble Cedex 9,
France
\and Max-Planck-Institut f\"ur Chemische Physik fester Stoffe, 01187 Dresden, Germany
}

\date{Published in European Physical Journal B {\bf 21}, 535 (2001)}

\abstract{
We report results of a new analysis for the low-temperature structure of
$\alpha$'-NaV$_{2}$O$_{5}$ from synchrotron x-ray diffraction experiments. We
confirm the existence of two inequivalent ladder structures in each vanadium
layer. 
Based on our structural data we
perform a bond-valence calculation for the vanadium sites in the low
temperature state. Due
to an asymmetric charge ordering we obtain only two different vanadium valences
despite the three inequivalent sites. This explains the $^{51}$V-NMR
observation of only two resonant peaks in the charge ordered phase.
By use of a Slater-Koster method to obtain hopping matrix elements and cluster
calculations we obtain effective vanadium-vanadium hoppings which compare well
to LDA results. Using these in a cluster calculation we obtain a superexchange
of $0.047$ eV between electrons on neighbouring rungs of the same ladder for
the undistorted phase. For the distorted phase we find a significant
alternation in the shifts of the oxygen atoms along the legs of one of the two
ladder types which leads
to a significant exchange dimerisation $\delta_{J}\approx 0.25$. 
} 


\maketitle


\section{Introduction}
\label{intro}

The layered oxide $\alpha$'-NaV$_{2}$O$_{5}$ has attracted great interest
since 1996, when Isobe and Ueda reported a phase transition at $T=34$K with a
spin-Peierls like spin gap formation~\cite {isobe96}. An earlier determination
of the high-temperature structure~\cite {carpy75} originally reported two
inequivalent vanadium sites. However, a redetermination of the crystal
structure of the high-temperature phase found the space group Pmmn 
\cite {schnering98,meetsma98,smolinski98} instead of P2$_1$mn, with only one
type of vanadium site. Due to stoichiometry this site 
should have a valence of $+4.5$. Later, a $^{51}$V-NMR study by Ohama et
al~\cite {ohama98} reported the existence of two inequivalent vanadium sites
below the critical temperature. This is supposed to be caused by a charge
ordering transition of the material which leads to inequivalent
vanadium atoms with different valences. A determination of the
low-temperature structure by x-ray scattering then 
reported~\cite {ludecke99}\textit {three} inequivalent vanadium sites. Using
this data a bond-valence calculation then led to three different
valences,~\cite {smaalen00} approximately given by $+4$, $+4.5$, and $+5$,
in contrast to the result of the $^{51}$V-NMR measurements.

In this article, we will give the results of a
determination of the low-temperature structure using a much larger data set 
than used in~\cite{ludecke99}. 
Applying bond-valence analysis to our data we obtain only
\textit {two} significantly different valences for the vanadium sites,
despite the fact that our structural data show also three inequivalent sites. 

Due to shifts in the oxygen positions we further find an alternation in
the exchange coupling constant between 
the vanadium sites on one type of ladder and give an estimate for the
corresponding exchange dimerisation parameter $\delta_{J}$.

This article is organised as follows. In the next section we present the
experimental details and the results of the structure determination. We find
three inequivalent vanadium sites in each layer, building two inequivalent
vanadium-oxygen ladders per layer. In the
third section we use the bond-valence method to determine the valences of
the sodium and the vanadium atoms both in the undistorted and the distorted
phase and compare the results with the observations of NMR. In the fourth
section we look at the superexchange coupling along ladder direction for
the undistorted phase and for the two ladders per layer of the distorted
phase. To do this we first determine approximate vanadium-oxygen and
sodium-oxygen hopping matrix elements. From these we find the effective
vanadium-vanadium hopping elements by use of a block diagonalization comparing
our results with those of a recent LDA calculation~\cite{yaresko00}. We
proceed to find the superexchange coupling for the different ladder types and
determine its alternation along the ladder direction. In the last section we
summarize and discuss our results.

\begin {table*} [tbp]
\caption {Average structure parameters obtained by using fundamental
reflections at $15$K.}
\label {hight}
\begin {tabular} {llllllll}
\hline\noalign{\smallskip}
Site & \multicolumn{1}{c}{x} &  \multicolumn{1}{c}{y}
&  \multicolumn{1}{c}{z} & Site & \multicolumn{1}{c}{x} &
\multicolumn{1}{c}{y}
&  \multicolumn{1}{c}{z}\\
\noalign{\smallskip}\hline\noalign{\smallskip}
V & 0.40212(7) & 0.25000(0) & 0.39067(22) &
Na & 0.25000(0) & -0.25000(0) & 0.85362(73)\\
O$_{1}$ & 0.25000(0) & 0.25000(0) & 0.51111(122) &
O$_{2}$ & 0.38565(42) & 0.25000(0) & 1.05413(151)\\
O$_{3}$ & 0.42595(39) & -0.25000(0) & 0.50381(93) &
Cell~\cite{nakao98} & 11.3030(1)\AA & 3.61095(3)\AA & 4.7525(1)\AA\\
\noalign{\smallskip}\hline
\end {tabular}
\end {table*}

\section {Structure details}
\label{sec:1}

The crystal structure of the room temperature phase of NaV$_2$O$_{5}$
based on X-ray diffraction intensity data has been reported previously
\cite{schnering98}.
We have measured the diffraction intensities of
NaV$_2$O$_{5}$ at T = 15 K in the charge ordered phase using the same crystal.
The experiment
was performed using a single crystal of size $0.03
\times 0.09 \times 0.20$ mm$^{3}$ on the
diffractometer D3 at the synchrotron X-ray source of the HASYLAB at
DESY, Hamburg by using X-ray wavelength $\lambda = 0.559${\AA}.
The sample was mounted inside a
Displex cooling device attached to the four-circle diffractometer.
The superlattice reflections were detected corresponding to the
lattice parameters $a = 2\times a_{0}$, $b = 2\times b_{0}$ and $c = 4
\times c_{0}$ where $a_{0}$, $b_{0}$ and $c_{0}$ are subcell lattice
parameters. To
determine the space group, reflections were measured using the primitive super
cell for small values of Bragg angles which showed that the space group is
$Fmm2$.  4847 reflections were measured at T$=15$K up to
$\sin{\theta}/\lambda = 0.94${\AA}$^{-1}$ which gave rise to
1953 unique reflections. We have determined the average structure by
using only the fundamental reflections and the room temperature
structure model in space group $Pmmn$.
Least-squares refinement of the crystal
structure by using SHLEX program led to the atom parameters listed in
table~\ref{hight}. The residual index R, defined as
\begin {equation}
R=\frac{\sum||F_{0}|-|F_{c}||}{\sum|F_{0}|}
\end{equation}
was $R=0.060$. The average structure obtained agrees closely with the
room temperature structure\cite{schnering98} as expected.
We next refined the charge-ordered distorted low temperature
structure in the space group $Fmm2$. The number of parameters refined
were 61. The starting parameters were the parameters of the average
structure (table~\ref{hight})
transformed appropriately to the supercell.
The resulting atom coordinates have been given in table~\ref{lowt}.
Note that contrary to the superspace
group refinement of reference~\cite{ludecke99} we have employed the standard
crystallographic refinement procedure which we could afford due to the
very large number of reflection intensities measured up to a large
value of  $\sin{\theta}/\lambda = 0.94${\AA}$^{-1}$.
Accordingly our parameters were free from the constraints of the
refinement using the superspace group. Our refinement also led to
physically meaningful thermal parameters. The
residual index was R$=0.056$. The atom coordinates are
listed in table~\ref{lowt}.
To ensure that the refinement did not lead to a local minimum we used
the so-called multistart-and-refine procedure in which one starts
from several initial displaced coordinates. These refinements lead to
identical coordinates within the standard deviations. Recently
doubts have been raised about the the actual space group of
NaV$_2$O$_{5}$. $^{23}$Na-NMR results have suggested a lower-symmetry
space group than $Fmm2$~\cite{ohama00}. 
We did not attempt to refine the crystal
structure by lowering the space group symmetry
because of the very large number of parameters and consequent strong
correlations among the parameters involved. Also there
exists no diffraction evidence for a lower symmetry space group.
Details of the present
refinement procedure will be published elsewhere~\cite{tchatterji2be}.

\begin {table*} [tbp]
\caption {Low temperature phase structure at T = 15 K and shifts
from high temperature
structure as given in~\cite{schnering98}. Size of unit cell taken
from~\cite{nakao98}. Shifts from the positions in~\cite{schnering98} have been
calculated after shifting the low temperature unit cell by the vector
(-1/8, 1/8, 0.87959/4). The shift in z direction is chosen, such that the
positional change of V2 and V4 is nearly zero and symmetrical.
Thus, the vanadium sites of the spin
dimerised ladders do not move in z direction below T$_C$.}
\label {lowt}
\begin {tabular} {lllllll}
\hline\noalign{\smallskip}
&\multicolumn {3}{c}{Coordinates} &\multicolumn {3}{c}{Shift (new-old)}\\
Site & \multicolumn{1}{c}{x} & \multicolumn{1}{c}{y} & \multicolumn{1}{c}{z} &
\multicolumn{1}{c}{$\Delta$x} & \multicolumn{1}{c}{$\Delta$y} &
\multicolumn{1}{c}{$\Delta$z}\\
\noalign{\smallskip}\hline\noalign{\smallskip}
V1 (V$_{21a}$) & 0.32777(15) & 0.00000(0) & -0.06039(19) & 0.00172 &
0.00000 & -0.00336\\
V2 (V$_{1a}$)& 0.07595(6) & 0.24992(3) & 0.49682(8) & -0.00011 & -0.00008 &
-0.00001\\
V3 (V$_{22b}$)& 0.32460(16) & 0.00000(0) & 0.19398(23) & -0.00146 & 0.00000
& 0.00102\\
V4 (V$_{1b}$)& 0.07611(6) & 0.25013(3) & 0.24684(8) & 0.00006 & 0.00013 &
0.00001\\
V5 (V$_{22a}$)& 0.32469(17) & 0.00000(0) & 0.44451(19) & -0.00137 & 0.00000
& 0.00155\\
V6 (V$_{21b}$)& 0.32771(16) & 0.00000(0) & 0.69017(24) & 0.00166 & 0.00000
& -0.00280\\
Na1 & 0.25000(0) & -0.25000(0) & -0.19233(32) & 0.00000 & 0.00000 & -0.00205\\
Na2 & 0.25000(0) & -0.25000(0) & 0.05830(32)& 0.00000 & 0.00000 & -0.00142\\
Na3 & 0.00000(0) & 0.00000(0) & 0.13004(30)& 0.00000 & 0.00000 & -0.00003\\
Na4 & 0.00000(0) & 0.00000(0) & 0.38016(29)& 0.00000 & 0.00000 & 0.00009\\
Na5 & 0.00000(0) & 0.50000(0) & 0.37925(30)& 0.00000 & 0.00000 & -0.00082\\
Na6 & 0.00000(0) & 0.50000(0) & 0.12955(31)& 0.00000 & 0.00000 & -0.00052\\
O1 & 0.25133(11) & 0.00000(0) & 0.22409(48) & -0.00133 & 0.00000 & 0.00053\\
O2 & 0.24930(15) & 0.00000(0) & -0.02628(52) & -0.00070 & 0.00000 & -0.00090\\
O3 & 0.00000(0) & 0.25299(34) & 0.46549(50) & 0.00000 & 0.00299 & 0.00032\\
O4 & 0.00000(0) & 0.25283(35) & 0.21551(50) & 0.00000 & 0.00283 & 0.00034\\
O5 & 0.08905(29) & 0.50000(0) & 0.21785(35) & 0.00050 & 0.00000 & 0.00098\\
O6 & 0.08763(28) & 0.00000(0) & 0.46609(33) & -0.00092 & 0.00000 & -0.00078\\
O7 & 0.33823(28) & 0.24747(20) & 0.22238(31) & -0.00032 & -0.00253 & -0.00054\\
O8 & 0.33865(29) & -0.25153(28) & -0.02744(31) & 0.00010 & -0.00153 &
-0.00036\\
O9 & 0.08746(28) & 0.00000(0) & 0.21637(34) & -0.00109 & 0.00000 & -0.00050\\
O10 & 0.08914(29) & 0.50000(0) & 0.46812(33) & 0.00059 & 0.00000 & 0.00125\\
O11 & 0.31682(28) & 0.00000(0) & 0.10863(37) & -0.00103 & 0.00000 & -0.00072\\
O12 & 0.31636(27) & 0.00000(0) & -0.14520(31) & -0.00149 & 0.00000 & -0.00455\\
O13 & 0.31669(28) & 0.00000(0) & 0.60519(35) & -0.00116 & 0.00000 & -0.00416\\
O14 & 0.31686(27) & 0.00000(0) & 0.35913(32) & -0.00099 & 0.00000 & -0.00022\\
O15 & 0.06871(29) & 0.24749(18) & 0.58013(31) & 0.00086 & -0.00251 & -0.00032\\
O16 & 0.06979(29) & 0.24890(27) & 0.33107(34) & 0.00194 & -0.00110 & 0.00062\\
Cell & 2$\times$11.3030(1){\AA} & 2$\times$3.61095(3){\AA} &
4$\times$4.7525(2){\AA}\\
\noalign{\smallskip}\hline
\end{tabular}
\end {table*}

\begin {figure*} [tbp]
\includegraphics{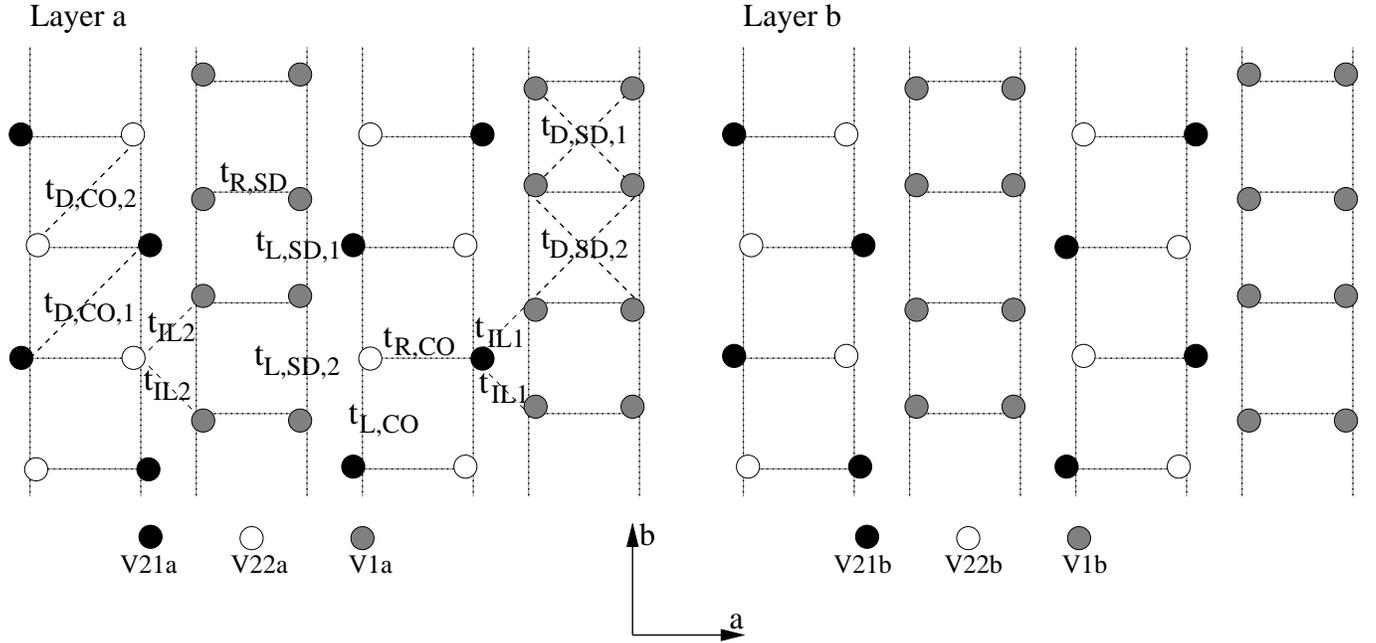}
\caption {Vanadium layer of the low-temperature
structure. Note the difference in the shifts of V$_{1a}$ and V$_{1b}$.
Figure not to scale.}
\label {vanadiumladders}
\end {figure*}

We get the same qualitative results as were found
in~\cite{ludecke99,deBoer00}.  
Two slightly different types of vanadium layers were found. Both layers
look similar as shown in figure~\ref{vanadiumladders} with three inequivalent
sites and as described in~\cite{ludecke99}. We denote the two different layers
by the subscripts a and b and the vanadium sites as shown in
figure~\ref{vanadiumladders}. Apart from the slightly different magnitude of
the shifts the main qualitative difference is, that for layer a the V$_{1a}$
sites shift in b direction towards the V$_{21a}$ sites while for layer b the
V$_{1b}$ sites are found to shift away in b-direction from the V$_{21b}$
sites. 

\section {Bond-valence calculation}
\label{sec:2}

The bond-valence method serves to determine the valences of atoms in a
chemical compound~\cite {brown73}. The valence of an atom in a compound, as
determined by this method, is the sum of the bond valences for all bonds, in
which the atom participates. The bond valences of such anion-cation bonds are
empirically defined as 
\begin {equation}\label {bvmethodb}
v_{i}=\exp \left[\left(r_{0}-r_{i}\right)/B\right],
\end {equation}
where $B$ is fixed to be $0.37$. $r_i$ is the distance between the two atoms
of the anion-cation bond, $r_0$ is an empirical parameter obtained from
fitting the atomic valences for many different compunds at room
temperature~\cite{brown85}. There is an alternative definition for the
bond valence, namely
\begin {equation}
v_{i}=\left(\frac{r_i}{r_0}\right)^{-N},
\end{equation}
where both $r_0$ and $N$ are empirical parameters. Both definitions give
similar results; we will adopt definition (\ref{bvmethodb}) throughout this
article, as this method was also used in~\cite{smaalen00}.

It should be noted that the bond valence model does not imply a certain
physical nature of the bond. It works both with covalent and ionic type
bonds~\cite{brown92}. 
Inaccuracies generally come from two sources: First, the errors of
the structure determination and the lattice constant determination. Second,
the inaccuracy of the parameter $r_0$ in (\ref{bvmethodb}). This parameter is
determined from fitting it to a large number of structures containing a given
bond type and therefore carries an error with it. 
We give the total standard deviation
derived from these two sources of errors in our results for the valences. In
our calculations for the vanadium site valences the inaccuracies from the
structure determination and the lattice constant play the leading role. 

Therefore, 
in a first step we use the method to determine the valences of the atoms in
the undistorted phase. This enables us to adjust parameters to account for the
peculiarities of the material as well as to estimate the accuracy of the
method. In a second step we determine the valences of the vanadium and sodium
atoms in the distorted phase. This will give us some insight into likely
charge order structures.

\subsection {Atomic valences in the undistorted phase}
\label{sec:2.1}

We have obtained the bond lengths for the vanadium-oxygen bonds and the
sodium-oxygen bonds from the
structural data in~\cite{schnering98} and give them in tables~\ref
{vanox1} and~\ref{naox1}.

\begin {figure} [tbp]
\includegraphics{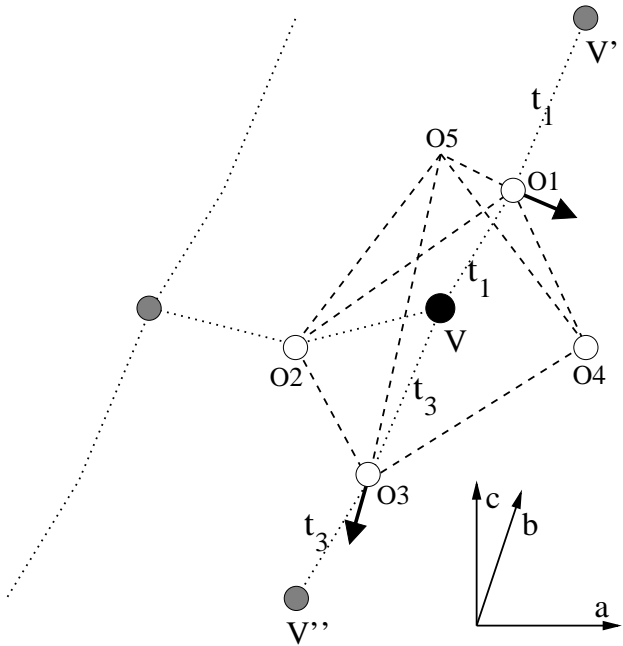}
\caption {Oxygen (white) and vanadium atoms (grey) around vanadium atom
(black) on same ladder. Arrows indicate shift of oxygen atoms along the leg in
the V$_{1}$-ladders. Figure not to scale.}
\label {oxygenpyramid}
\end {figure}

\begin {table}[tbp] 
\caption {Vanadium-oxygen bond lengths given in {\AA}ngstr{\o}m. V$_0$ denotes
the vanadium site of the undistorted phase. For the nomenclature of the
oxygen sites, see figure~\ref{oxygenpyramid}. The first row gives the distances for
room temperature lattice parameters as given in~\cite{schnering98}, the second
row gives distances for lattice parameters at 15K --- see table~\ref{lowt}.}
\label {vanox1}
\begin {tabular} {llllll}
\hline\noalign{\smallskip}
& O$_{1}$ & O$_{2}$ & O$_{3}$ & O$_{4}$ & O$_{5}$\\
\noalign{\smallskip}\hline\noalign{\smallskip}
V$_{0}$   & 1.916(3) & 1.825 & 1.916(3) & 1.986(3) & 1.616(3)\\
V$_{0}$   & 1.914(3) & 1.8216(7) & 1.914(3) & 1.983(3) & 1.600(3)\\
\noalign{\smallskip}\hline
\end {tabular}
\end {table}

\begin {table}[tbp] 
\caption {Sodium-oxygen bond lengths given in {\AA}ngstr{\o}m for room
temperature lattice constants (first row) and lattice constants at 15K (second
row). Na0 denotes the
sodium site of the undistorted phase; for the nomenclature of the oxygen sites
see figure~\ref{naov-nomenclature}. Due to symmetry there are always pairs of
bonds with the same length, thus, the lengths of only four bonds are given.}
\label {naox1}
\begin {tabular} {lllll}
\hline\noalign{\smallskip}
& O$_{R}$ & O$_{L}$ & O$_{A1}$ & O$_{A2}$ \\
\noalign{\smallskip}\hline\noalign{\smallskip}
Na0 & 2.435 & 2.606 & 2.554 & 2.554\\
Na0 & 2.424 & 2.594 & 2.550 & 2.550\\ 
\noalign{\smallskip}\hline
\end {tabular}
\end {table}

\begin {figure} [tbp]
\includegraphics{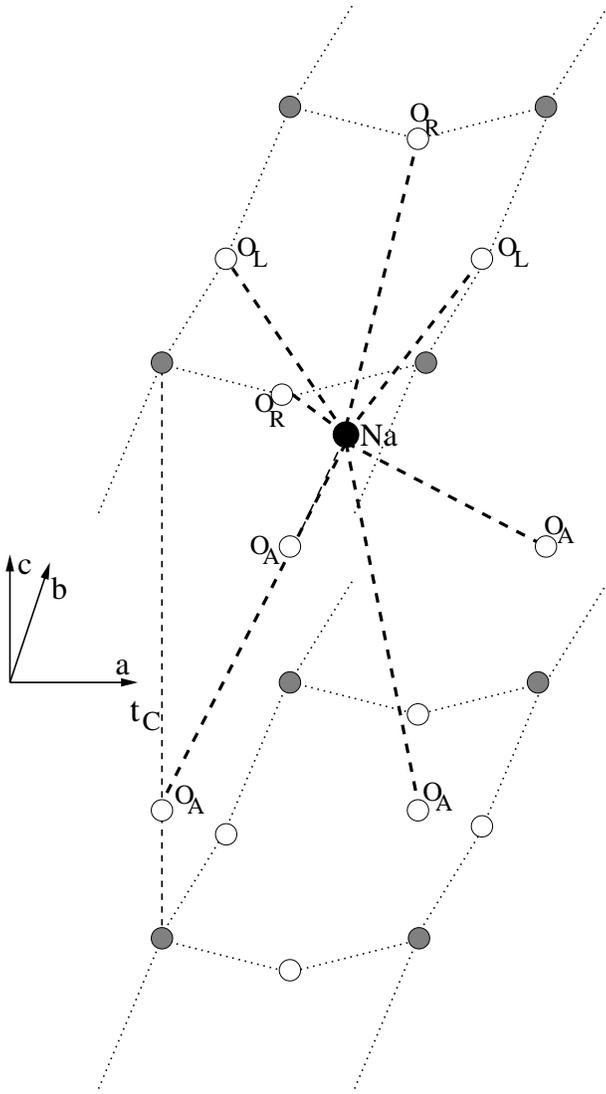}
\caption {Sodium (black), oxygen(white) and vanadium (grey) atoms on two
ladders in two neighbouring layers. Figure not to scale.}
\label {naov-nomenclature}
\end {figure}

Using the lengths of the vanadium-oxygen bonds we determine the valences of
the vanadium and the sodium atoms in the undistorted phase
by~\ref{bvmethodb}. Using room temperature lattice parameters and the
empirical $r_0$ given in~\cite{brown85} we obtain two different results,
depending on which value we take for r$_0$. Using the value for V$^{4+}$--O
bonds  r$_{4+}=1.784(27)${\AA} gives a valence of 4.45; using the value for
V$^{5+}$--O bonds r$_{5+}=1.803(31)${\AA} leads to 
a valence of 4.69. We know from
stoichiometry that the vanadium atom should have a valence of
$4.5$. Just using the average of the two r$_0$ given above is not a good
approximation, since the characteristic bond lengths do not vary linearly with
the bond order. Therefore, we use a fitted r$_{4.5+}=1.788(5)${\AA} to give the
correct 
valence 4.5 for the undistorted phase with room temperature lattice
parameters. 

The size of the unit cell decreases significantly from 300K to 15K. We
therefore cannot simply use the  r$_{4+}$,  r$_{4.5+}$ and  r$_{5+}$ from
above, since they result from room temperature structure determinations. Using
the high temperature r$_{4.5+}$ with the low temperature lattice parameters
for determination of the valence of the vanadium site in the undistorted
phase would give a valence of 4.59. We therefore have to scale down all
these characteristic bond lengths. For this scale we use the factor
$(\mbox{V}_{{15K}}/\mbox{V}_{{300K}})^{1/3}=0.9965$ where
the V are the volumes of a cell containing four vanadium sites. We then
obtain a valence of 4.51 for the vanadium atom in the undistorted phase
using 
low temperature lattice constants and the reduced r$_{4.5+}$.

We can follow through a similar exercise for the sodium atom. We find that the
characteristic bond length $r_{Na}=1.803${\AA} given in~\cite{brown85} gives a
valence of 1.12 for the sodium atom in the undistorted phase with room
temperature lattice parameters, which is clearly deviating strongly from the
expected value 1. It is known, however, that there is a comparatively strong
physical variation in the environment of alkali atoms, therefore the
characteristic bond lengths vary strongly, the standard deviation being 
about 0.08{\AA}~\cite{brown85}. It is therefore
justified to use a characteristic bond length $r_{Na}=1.762(5)${\AA} obtained
by fitting the resulting valence to 1. This helps capturing the special
physical features of the environment surrounding the alkali atom in this
material. Scaling down this characteristic $r_{Na}$ to $1.756(5)$ for lattice
constants at 15 K as in the
case of 
vanadium and using these lattice constants we arrive at a
valence of $1.00(1)$ for the sodium atom in the undistorted phase using low
temperature lattice parameters. Since $r_{Na}$ is a fitted value, the error
comes only from the standard deviation of the structure determination and is
much smaller than the standard deviation of $r_{Na}$ from~\cite{brown85}.
All the parameters used are listed in
table~\ref{bvparameter}. Henceforth we will always use the lattice parameters
and characteristic bond lengths for temperature 15K as given there.

\begin {table}[tbp] 
\caption {Parameters used for the determination of atomic valence. Lattice
constants from~\cite{schnering98,nakao98}, for other parameters see text.}
\label {bvparameter}
\begin {tabular} {lll}
\hline\noalign{\smallskip}
& T$=300$K & T$=15$K\\
\noalign{\smallskip}\hline\noalign{\smallskip}
a & 11.311 {\AA} & 11.3030 {\AA}\\
b & 3.6105 {\AA} & 3.61095 {\AA}\\
c & 4.800 {\AA} & 4.7525 {\AA}\\
r$_{4+}$ & 1.784(5) {\AA} & 1.778(27) {\AA}\\
r$_{4.5+}$ & 1.788(5) {\AA} & 1.782(5) {\AA}\\
r$_{5+}$ & 1.803(3) {\AA} & 1.797(31) {\AA}\\
r$_{Na}$ & 1.762(5) {\AA} & 1.756(5) {\AA}\\
V & 196.0 {\AA}$^3$ & 194.0 {\AA}$^3$\\
\noalign{\smallskip}\hline
\end {tabular}
\end {table}

To check the quality of the bond valence method we now determine the valence
of the three oxygen sites of the undistorted phase. Since the bond orders are
already determined, the total valence of the five oxygen atoms in a unit cell
is $-10.02(6)$, corresponding to the total valence of the cations. However the
valence 
of the different oxygens is not the nominal value $-2$ for 
each as might be expected. Instead we
obtain a valence of $-1.87(2)$ for the apical oxygen, $-2.08(2)$ for the
oxygen on 
the leg of the V-O-ladder, and $-2.12(2)$ for the oxygen on a rung. 
Qualitatively this may be
explained with the fact that the oxygen atoms are easily polarizable.

\subsection {Atomic valences in the distorted phase}
\label{sec:2.2}

We will now determine the valence of the vanadium and the sodium atoms in the
distorted phase at 15K. For this we will use parameters as given in
table~\ref{bvparameter}. 

The bond lengths of the vanadium-oxygen bonds and of the sodium-oxygen bonds
as well as the valences resulting 
are listed in tables~\ref{vanox2} and~\ref{naox2}. Due to the
fact that the exact valence of the vanadium atoms is not known a priori and
that the characteristic bond length r$_0$ depends on this valence, one has to
use a self consistency procedure to derive the valences for the vanadium
atoms as given in the last column of table~\ref{vanox2}. 
\begin{table*}[tbp]
\caption {Vanadium-oxygen bond lengths given in {\AA}ngstr{\o}m and valences
in the distorted phase at 15K. Nomenclature for atoms as given in
figures~\ref{oxygenpyramid} and~\ref{vanadiumladders}.
For the valences The first three columns show
the valence depending on the characteristic bond length r$_0$ used, the last
row gives values which are self-consistent to first order.}

\suppressfloats

\label {vanox2}
\begin {tabular} {llllllllll}
\hline\noalign{\smallskip}
&&&&&&\multicolumn{3}{c}{Valences}\\
& O$_{1}$ & O$_{2}$ & O$_{3}$ & O$_{4}$ & O$_{5}$&r$_{4+}$&r$_{4.5+}$ &r$_{5+}$ &SC \\
\noalign{\smallskip}\hline\noalign{\smallskip}
V$_{1a}$  & 1.910(2) & 1.817(4) & 1.915(3) & 1.985(7) & 1.592(6)&
 4.51(4) & 4.56(4) & 4.75(4)& 4.60(4) \\
V$_{21a}$ & 1.937(4) & 1.889(8) & 1.937(4) & 1.978(7) & 1.633(5)&
 4.11(4) & 4.15(4) & 4.32(4) & 4.12(4)\\
V$_{22a}$ & 1.898(3) & 1.762(9) & 1.898(3) & 1.999(7) & 1.633(5)&
 4.52(4) & 4.57(4) & 4.76(4) &4.61(4)\\
V$_{1b}$  & 1.914(3) & 1.821(4) & 1.909(3) & 1.992(7) & 1.608(7)& 
4.43(4) & 4.48(4) &4.66(4) &  4.48(4)\\
V$_{21b}$ & 1.938(3) & 1.900(7) & 1.938(3) & 1.981(7) & 1.635(6)&
 4.07(4) & 4.11(4) & 4.28(4) & 4.07(4)\\
V$_{22b}$ & 1.892(3) & 1.752(7) & 1.892(3) & 2.004(7) & 1.632(6)&
 4.57(4)&4.62(4)& 4.81(4)&  4.69(4)\\
\noalign{\smallskip}\hline
\end {tabular}
\end {table*}

\begin {table}[tbp] 
\caption {Sodium-oxygen bond lengths  given in
{\AA}ngstr{\o}m and valence in the distorted phase at 15K. Nomenclature for
oxygen atoms as shown in figure~\ref{naov-nomenclature}. 
Due to symmetry there are
pairs of bonds of equal length, only one of each pair is given.}
\label {naox2}
\begin {tabular} {llllll}
\hline\noalign{\smallskip}
& O$_{R}$ & O$_{L}$ & O$_{A1}$ & O$_{A2}$ & Valence\\
\noalign{\smallskip}\hline\noalign{\smallskip}
Na1 & 2.405 & 2.570 & 2.513 & 2.550&1.06(8)\\
Na2 & 2.418 & 2.583 & 2.541 & 2.515 & 1.04(8) \\
Na3 & 2.444 & 2.569 & 2.576 & 2.576 & 0.97(8) \\
Na4 & 2.443 & 2.568 & 2.567 & 2.567 & 0.98(8) \\
Na5 & 2.423 & 2.630 & 2.572 & 2.572 & 0.96(8) \\
Na6 & 2.420 & 2.621 & 2.548 & 2.548 & 1.00(8)\\
\noalign{\smallskip}\hline
\end {tabular}
\end {table}

Regarding the sodium atom valence one observes, that the changes of valence
are smaller than the standard deviation, which is mostly due to the standard
deviation of r$_{Na}$. Furthermore the average valence does not change
significantly, it is equal to $1.01(3)$. 

Regarding vanadium, the first observation is that the valences of similar
vanadium atoms in the two layers are not the same within one standard
deviation. We will therefore look at these layers separately. 

We first look at the valences for layer $a$. Although the V$_{1a}$ and the 
V$_{22a}$ sites are crystallographically inequivalent they 
have approximately the same valence $4.6$. 
The V$_{21a}$ site has a lower valence of approximately $4.1$. Note that the
shift from the average valence 4.5 is significantly different for V$_{21a}$
and V$_{22a}$.

For layer $b$ the results are somewhat different. Here the V$_{1b}$ site
experiences only a small change of valence. Despite this the absolute size of
the valence shifts
from the original average valence 4.5, which are $0.43(4)$ for V$_{21b}$ and
$0.19(4)$ for V$_{22b}$, are significantly different from each other.

To determine the quality of the results, one can calculate the
stoichiometrically averaged valence of a vanadium site for the low-temperature
structure. Due to charge
conservation one would ideally expect an average valence of $4.5$.
Using the self-consistently interpolated values one obtains 
$4.48(2)$ for layer $a$ and $4.43(2)$ for layer $b$. 
This indicates that the results for layer $a$ should be the more reliable. The
difference between the results for the different layers can well be caused by
the refinement method of the structure determination, which might have found a
less reliable local minimum for layer $b$ than for layer $a$.
From these bond valence calculations for layer a we therefore 
conclude that there are  
\textit{only two} significantly different vanadium valences in the
distorted phase, contrary to the result of
van Smaalen and L\"udecke, who obtained three clearly different valences
\cite {smaalen00} and a nearly symmetrical charge shift between the 
V$_{21}$ and the V$_{22}$ sites. However, in our analysis we 
obtain a significantly asymmetrical charge shift 
between the V$_{21}$ sites on one hand and the V$_{1}$ and V$_{22}$ sites
on the other hand for both layers. 

Such a charge transfer between ladders does not have to cause the material to
become a conductor. The insulating behaviour of the high temperature phase has
been explained with correlation effects, due to which one can map the 
quarter filled ladder to a half-filled Hubbard chain~\cite{horsch98}, 
resulting in an
effective charge transfer gap. Since
this argument works only at quarter-filling, one might expect
the material to become a conductor if a charge transfer takes place between
ladders as described above. 
While this is true if one considers isolated ladders, the situation
becomes less simple if inter-ladder Coulomb repulsion is included in
the model, due to the geometry of the lattice. Consider a hexagon of 
V$_{1}$ and V$_{22}$ sites and assume that the V$_{21}$ sites carry one
electron each. In this case one has two electrons for each six other sites,
close to what is found from bond valence above. As a consequence, 
one still has a charge
transfer gap, because due to the geometry of
the lattice it is not possible to put three electrons on such a hexagon
without having Coulomb repulsion. Therefore depending on the ratio of the
parameters involved, the system can still be a charge transfer insulator.

\subsection {Comparison with NMR}
\label {sec2:3}

We can compare our result to the results of the $^{51}$V-NMR measurements
\cite {ohama98}. For such a measurement one expects that the resonant
peak for the high-temperature vanadium site with valence $+4.5$ splits into
multiple peaks below the transition: one peak for each different valence of
vanadium sites is expected. The intensity of a peak for a given valence should
be approximately proportional to the number of vanadium sites having such
valence. 

The interaction between the 3d electrons and the nuclei of the vanadium
which causes the shift of the NMR resonance frequency, is, disregarding
crystal field effects due to the distortion, roughly proportional to the
density of the $d_{xy}$-electrons on a shell of a given vanadium
ion. Therefore one expects that the splitting of the resonant NMR-peak at the
transition temperature reflects the charge disproportionation approximately
linearly, at least close to the critical temperature. 

For a low-temperature structure with three different valences $4.5-\delta$ for
the V$_{21}$ sites, $4.5$ for the V$_{1}$ sites, and $4.5+\delta$ for the
V$_{22}$ sites, as suggested by~\cite {smaalen00}, one expects a splitting of
the single high-temperature resonant peak into \emph {three} different peaks
with an intensity 
ratio of approximately $1:2:1$, the corresponding shifts of the right and left
peak should be roughly symmetrical around the center peak and the center peak
in continuation of the single high temperature resonance peak.
For a low-temperature structure with only two different valences
$4.5-3\delta$ for the V$_{21}$ sites, and $4.5+\delta$ for the V$_{22}$ and
the V$_{1}$ sites as suggested by our analysis above, one
expects a splitting of the high-temperature resonant peak into \emph {two}
peaks at the transition temperature with an intensity ratio of
approximately $1:3$ and where the shifts of the new peaks from the old peak
obey a $3:1$ ratio.

Looking at the experimental results as given in figure 3 of~\cite{ohama98},
one 
finds, that \emph {two} resonant peaks have been found below $33.4$K. The
shift of these two peaks relative to the high-temperature peak is clearly
asymmetric, 
and the ratio of the distances is roughly $2:1$ in favour of the left peak
which has been identified to be a V$^{4+}$ resonant peak by the authors. 
The ratio of the
intensities, as determined by the area below the peaks, is roughly $1:3$ from
$30$K to $33.3$K. This changes at $20$K, but it is not mentioned
in~\cite{ohama98} whether intensity correction has been performed for the
data, so as far as intensity ratios are concerned 
we use only the data around T$_C$. 

From $33.4$K to $33.6$K
three peaks are found. One of them is clearly the continuation of the
high-temperature peak for valence $4.5$. This behaviour might be intrinsic or
due to a distribution of the T$_C$ in the powder sample used for the
measurement. In the former case (intrinsic behaviour) one would identify the
center peak with the  
V$_{1}$ sites and the left and right peak with the V$_{21}$ and the V$_{22}$
sites, respectively. Since the charge of the V$_{1}$ sites would then remain
constant in this temperature interval, the charge shift of the V$_{21}$ and
the V$_{22}$ sites would be of the same amount and different sign. This
implies a ratio of roughly $1:1$ for the distances of the two side peaks from
the center peak, i.e., a different ratio than that measured below $33.4$K. 
For the latter case (distribution of T$_C$) one would not expect a big change
of the distance ratio at $33.4$K. Since one finds that 
the ratio of distances is still $2:1$ from $33.4$K to $33.6$K, we believe that
the observation of three peaks is due to a distribution of the 
T$_C$ rather than intrinsic. 
However, below $33.4$K the $^{51}$V-NMR measurements seem to be in much better
agreement with the results of our analysis, yielding \emph {only two}
different valences with an asymmetrical charge disproportionation, than with
the 
previous analysis of van Smaalen and L\"udecke yielding \emph {three}
different valences and symmetrical charge disproportionation~\cite
{smaalen00}. 
Our results also explain, that the low-valence V$_{21}$ sites can
have a valence of nearly $4$ as determined in~\cite {ohama98} from the
hyperfine coupling strengths, 
although the charge disproportionation $\delta_{\mbox{c}}$ 
in the system is estimated to be small: $\delta_{\mbox{c}}\ll 0.5$
~\cite {fagot99,sherman99}.

\section {Exchange coupling constant alternation}
\label{sec:3}

From the structural data we observe a slight alternation in
the distances in b-direction between V$_{1}$ sites for the V$_{1}$
ladder in both layers. The lattice dimerisation parameter $\delta_V$
for the shift of the V$_{1}$ sites in b direction is very 
small with $0.04\%$ for layer $a$ and $0.05\%$ for layer
$b$ compared to $0.56\%$ for CuGeO$_{3}$~\cite {hirota94}. However, even a
small alternation implies an alternation in the exchange coupling in
b-direction. 

Under the assumption that the observed spin-gap results from an alternation of
the exchange coupling in
\emph{all} ladders, the exchange dimerization parameter
$\delta_{\mbox{J}}=(\mbox{J}_{+}-\mbox{J}_{-})/(\mbox{J}_{+}+\mbox{J}_{-})$
is estimated to lie  between 
$3.4\%$ to $10\%$ for $\alpha$'-NaV$_{2}$O$_{5}$~\cite
{johnston00,konstantinovic99,lohmann97,vasilev97} and $3\%$
to $17\%$ for CuGeO$_{3}$~\cite {hirota94,hase93}. In both cases the higher
values result  
from mean-field estimates, which might be insufficient for a treatment of these
materials as argued for $\alpha$'-NaV$_{2}$O$_{5}$ in~\cite
{johnston00}. However, the assumption that there is an exchange coupling
alternation in all ladders of the distorted phase is flawed from the point of
structure determination, since we observe 
only for the V$_1$-ladders any alternation in b-direction, but not for the
V$_{21}$-V$_{22}$-ladders. 

Assuming that an alternation of the exchange coupling in b-direction takes
place only in every second ladder, a recent theoretical study estimated the
dimerisation parameter $\delta$ necessary to produce the observed size of the
spin gap as $\delta\approx 0.38\ldots 0.54$.~\cite{honecker00} This would be
even larger than the dimerisation observed in CuGeO$_{3}$.
Thus, it seems unlikely, that the shifts of the vanadium sites and the
resulting dimerisation in 
$\alpha$'-NaV$_{2}$O$_{5}$ which is an order of magnitude smaller than that
for CuGeO$_{3}$ should give an exchange dimerisation of the same
order of magnitude. 

However, the small dimerisation of the vanadium sites is not the only
alternation along the legs of the V$_{1}$ ladders. One finds a much stronger
alternation in 
the distances to the nearest oxygen atoms which are on the same leg as the 
V$_{1}$ atom in question. This alternation results from the shift of these
oxygen atoms in a- and c-direction. 
For the V$_{21}$-V$_{22}$ ladder no effective alternation
in b-direction is found, neither from shifts of the oxygen sites nor from
shifts of the vanadium sites.
 
It is possible to determine the exchange dimerisation parameter from the
structural data. Let us first give a very simple estimate. For
the superexchange interaction between rungs we have approximately
$J\propto t_{VV}^2$,
$t_{VV}\propto t_{VO}^2$, and, according to~\cite{harrison89},
$t_{VO}\propto d_{VO}^{-3.5}$. $t_{VO}$ is a hopping matrix element between the
vanadium $d_{xy}$-orbital and the p-orbital of the neighbouring oxygen on the
same rung. $t_{VV}$ is therefore an effective hopping matrix
element between neigbouring vanadium sites of on the same leg of a
ladder. This implies, that for small distortions which change the distance
$d_{VO}$ by $\pm\Delta d_{VO}$ we obtain 
$\delta_J=\Delta J/J\approx \pm 14 \Delta d_{VO}/d_{VO}$. 
For the vanadium-oxygen bond lengths between atoms on the
same leg of the V$_1$ ladder given in
table~\ref{vanox2}, one finds $\delta_J\approx 2\%$. This is
one order of magnitude larger compared to what one would have obtained by
using the shifts of the vanadium sites only. 

We will now first examine the hopping matrix elements for the undistorted
phase, calculating them with the method presented in~\cite{slater54} using the
parameters from~\cite{harrison89}. From this we will gain insight in the
effective $t_{VV}$ hopping matrix elements and the effective superexchange
between rungs on the same ladder. In a second step we extend this method onto
the distorted phase to find the approximate size of the superexchange
dimerisation on the V$_1$ ladders.

\subsection {Hopping and superexchange in the undistorted phase}
\label{sec:3.1}

\begin{table*}
\caption {Distances (in {\AA}ngstr{\o}m) between vanadium and oxygen atoms and
hopping matrix 
elements (in eV) between vanadium d$_{xy}$ and oxygen p-orbitals for the
undistorted 
phase. The absolute values of the hoppings and the distances are given, the
sign is determined by the orientation of the vector between the vanadium and
the oxygen site. Denomination of oxygen atoms as in figure~\ref{oxygenpyramid}. O$_{5'}$
denotes the apex oxygen of the VO$_5$ pyramid lying above or below the central
vanadium site.}
\label {vohopping1}
\begin {tabular} {llllllll}
\hline\noalign{\smallskip}
&$\Delta$x & $\Delta$y & $\Delta$z & d$_{VO}$ & t$_{p_xd_{xy}}$ & t$_{p_yd_{xy}}$ & t$_{p_zd_{xy}}$ \\
\noalign{\smallskip}\hline\noalign{\smallskip}
O$_{1}$, O$_{3}$ & 0.2825 & 1.8055 & -0.5695 & 1.914 & 0.8549 & 0.6303 & 0.2470\\
O$_{2}$ & 1.7193 & 0 & 0.6018 & 1.822 & 0 & 1.1632 & 0\\ 
O$_{4}$ & 1.9304 & 0 & 0.4545 & 1.983 & 0 & 0.8909 & 0\\
O$_{5}$ & 0.1855 & 0 & 1.5896 & 1.600 & 0 & 0.2247 & 0\\
O$_{5'}$ & 0.1855 & 0 & 3.1629 & 3.168 & 0 & 0.0104 & 0\\
\noalign{\smallskip}\hline
\end {tabular}
\end {table*}

\begin{table*}
\caption {Distances (in {\AA}ngstr{\o}m) between sodium and oxygen atoms and
hopping matrix 
elements (in eV) between sodium 3s and oxygen p-orbitals for the
undistorted 
phase. The absolute values of the hoppings and the distances are given, the
sign is determined by the orientation of the vector between the vanadium and
the oxygen site. Denomination of oxygen atoms as in figure~\ref{oxygenpyramid}.}
\label {naohopping1}
\begin {tabular} {llllllll}
\hline\noalign{\smallskip}
&$\Delta$x & $\Delta$y & $\Delta$z & d$_{NaO}$ & t$_{sp_x}$ & t$_{sp_y}$ & t$_{sp_z}$ \\
\noalign{\smallskip}\hline\noalign{\smallskip}
O$_{A1}$, O$_{A2}$ & 1.5338 & 1.8055 & 0.9434 & 2.550 & 1.2970 & 1.5267 & 0.7977\\
O$_L$ & 2.0018 & 0 & 1.6501 & 2.594 & 1.6076 & 0 & 1.3252\\
O$_R$ & 0 & 1.8055 & 1.6178 & 2.424 & 0 & 1.7768 & 1.5921\\
\noalign{\smallskip}\hline
\end {tabular}
\end {table*}

We determine the matrix elements for the electron hopping by a Slater-Koster
type of treatment presented in~\cite{harrison89} and~\cite{slater54}. 
We will use hopping matrix elements 
along anion-cation bonds and between vanadium orbitals only. For the
vanadium sites, we consider only the lowest lying 3d-orbital, which is the
d$_{xy}$ orbital according to~\cite{yaresko00}. For the oxygen sites we use
the 2p-orbitals, assuming them to be of equal energy, while 
for the sodium sites we
use the 3s-orbital. In this case, the hopping between vanadium and oxygen
sites is~\cite{slater54,harrison89}
\begin{eqnarray}
t_{p_xd_{xy}}=m\left(\sqrt{3}l^2 V_{pd\sigma}+
\left(1-2l^2\right)V_{pd\pi}\right) \nonumber\\
t_{p_yd_{xy}}=l\left(\sqrt{3}m^2 V_{pd\sigma}+
\left(1-2m^2\right)V_{pd\pi}\right) \\
t_{p_zd_{xy}}=lmn\left(\sqrt{3} V_{pd\sigma}- 2V_{pd\pi}\right) \nonumber 
\end{eqnarray}
\noindent
with $l$, $m$, $n$ being the direction cosines of the vector from the p
state to the d state and the parameters
\begin{eqnarray}
V_{pd\sigma}&=&-21.808 \textrm{ eV{\AA}}^{7/2} d_{VO}^{-7/2},\nonumber \\ 
V_{pd\pi}&=&10.054\textrm{ eV{\AA}}^{7/2} d_{VO}^{-7/2}.\nonumber
\end{eqnarray}
\noindent
The matrix elements for the Na-O hopping are determined from
\begin{eqnarray}
t_{sp_x}&=&lV_{sp\sigma}, \nonumber \\
t_{sp_y}&=&mV_{sp\sigma},\nonumber \\
t_{sp_z}&=&nV_{sp\sigma},\nonumber \\ 
V_{sp\sigma}&=&14.0208 \textrm{ eV{\AA}}^{2}d_{NaO}^{-2}.\nonumber
\end{eqnarray}
\noindent
Finally we may also consider hopping between nearest neighbour vanadium sites
on different rungs. In this case the matrix elements are given by
\begin{equation}
t_{d_{xy}d_{xy}}=3l^2m^2V_{dd\sigma} +\left(l^2+m^2-4l^2m^2\right)V_{dd\pi}
\end{equation}
\noindent
with parameters
\begin{eqnarray}
V_{dd\sigma}&=&-116.18 \textrm{ eV{\AA}}^{5} d_{VO}^{-5},\nonumber\\
V_{dd\pi}&=&62.754 \textrm{ eV{\AA}}^{5} d_{VO}^{-5}.\nonumber
\end{eqnarray}
\noindent
The coordinate differences, distances and resulting hopping elements between
the sites are given in tables~\ref{vohopping1} and~\ref{naohopping1}. For the
direct d$_{xy}$-d$_{xy}$ hopping matrix element 
one obtains $0.2231$ eV in the undistorted phase.

Using these matrix elements, we will now proceed to determine effective
vanadium-vanadium hopping in the following way. Consider the single particle
Hamiltonian $H_{SP}$ of the problem. In matrix notation it has the form
\begin{equation}
{\bf H}_{SP}=
\left(\begin{array} {lll}
{\bf T}_{VV} &{\bf T}_{VO}&{\bf T}_{VNa}  \\
{\bf T}_{OV} &{\bf T}_{OO}&{\bf T}_{ONa}  \\
{\bf T}_{NaV} &{\bf T}_{NaO}&{\bf T}_{NaNa}  
\end{array}\right).
\end{equation}
\noindent
The on-site energies $\epsilon_V$,
$\epsilon_O$, and $\epsilon_{Na}$ on the diagonal of ${\bf T}_{VV}$, ${\bf 
T}_{OO}$, and ${\bf T}_{NaNa}$ are given
in~\cite{harrison89} as  $\epsilon_V=-12.55$ eV, $\epsilon_O=-14.13$ eV, 
$\epsilon_{Na}=-5.13$ eV. One observes, that the hopping from one layer to
another via the intermediate apex oxygen is very small. However, this is not
so clear for hopping via the path V-O$_1$-Na-O$_5$-V. Therefore to find
effective vanadium-vanadium hopping matrix elements
we block-diagonalize ${\bf H}$ given
above with successive Jacobi transformations, projecting out 
the sub-matrices ${\bf T}_{VO}$, ${\bf T}_{NaO}$ and their
conjugates to separate the Hilbert space of the vanadium orbitals from that
of the rest. We then proceed to use these effective vanadium-vanadium
hoppings as parameters of an extended Hubbard model describing the system.

The block diagonalization is done for a cluster of 4 layers by 4 rungs by 4
ladders. The resulting values for the effective hoppings can be found in
table~\ref{vvhop1} and we can compare them with the results from an LDA
tight-binding fit presented in~\cite{yaresko00}. The results are partly
similar: t$_R$ 
is larger than both t$_L$  and t$_D$, although all these values are smaller
than those from LDA. t$_L$ and t$_R$ are of similar size.
t$_c$, the effective hopping between layers, is much smaller than all other
effective hoppings in both cases. 

\begin{table*}
\caption {Effective vanadium-vanadium hoppings for the undistorted phase given
in eV. Denomination of hoppings as given in figure~\ref{vanadiumladders}. For
methods see text. Superexchange J$_{eff}$ has been calculated with $U=4$ eV,
and $V_L=2t_R$ as given (see text).}
\label {vvhop1}
\begin {tabular} {llllllll}
\hline\noalign{\smallskip}
Method & t$_{R}$ & t$_{L}$ & t$_{D}$ & t$_{IL}$ & t$_{c}$ & V$_L$ & J$_{eff}$\\
\noalign{\smallskip}\hline\noalign{\smallskip}
LDA~\cite{yaresko00} & 0.38 & 0.085 & 0.085 & --- & --- & 0.76 & 0.070\\
Ab initio~\cite{suaud00} & 0.5382 & 0.1246 & --- & -0.0442 & --- & 1.076 & 0.038\\
All hoppings & 0.172 & 0.049 & 0.062 & -0.110 & 0.009 & 0.344 & 0.047\\
Without Na-O & 0.305 & 0.100 & 0.016 & -0.111 & $<0.001$ & 0.610 & 0.044\\
\noalign{\smallskip}\hline
\end {tabular}
\end {table*}

We can also compare our values with the results from an ab-initio
calculation using the DDCI2 quantum chemical method as presented
in~\cite{suaud00}. They obtain a t$_R$ which is larger than ours or that
from LDA calculation. They do not obtain a value for t$_D$ because
Na orbitals were not included in their
calculation. Therefore one should compare their value for t$_L$ with our
result for t$_D+$t$_L$. These values are similar. They also obtain a hopping
matrix element
t$_{IL}$ which is
smaller than ours by about a factor of 2.5. These differences are
probably due to the fact that they include the on-site repulsion energy
$U_p$ for the oxygen orbitals. This reduces the energy difference between
occupation of vanadium and oxygen orbitals and thus increases the effective
hopping between vanadium orbitals.

A strength of our method is that we are able to analyze 
the originof  particular matrix
elements. For example, regarding the
similar size of t$_D$ and t$_L$ we find, that it is due to the
inclusion of the Na-O hopping elements. Without these we get results as
given in the last row of table~\ref{vvhop1}, where t$_D$ is lower than t$_L$
by nearly two orders of magnitude. Similarly, we notice from
table~\ref{vanox1} that in our
model we have more vanadium-oxygen matrix elements than assumed in the
tight-binding fit 
in~\cite{yaresko00}. As we will see later, this is very important 
for the determination of the effect of small distortions. There one can no
longer use an effective one particle Hamilton operator where only a certain
subset of all hoppings is included.

We will next determine the effective superexchange J$_{eff}$ between electrons
on neighbouring rungs of the same ladder. We do this by considering the
Hamiltonian
\begin {eqnarray}
H&=&\sum_{i,j} \left(t^{VV}_{ij}a_i^\dag a_j + h.c. \right)+ U\sum_i
n_{i\uparrow}n_{i\downarrow}\nonumber \\
&&+\sum_{i,j} V_{ij}
\left(n_{i\uparrow}+n_{i\downarrow}\right)
\left(n_{j\uparrow}+n_{j\downarrow}\right),
\end{eqnarray}
\noindent
where the first term contains the effective hopping between vanadium d$_{xy}$
orbitals as obtained either from LDA or our method. The second term contains
the on-site Coulomb interaction U, the last term contains the
intersite Coulumb interaction V$_{ij}$ between sites $i$ and $j$. To determine
the effective superexchange we solve this Hamiltonian numerically on a cluster
of two neigburing rungs on the same ladder
with two d-electrons and set J$_{eff}$ to be the singlet-triplet
gap energy on this cluster. We disregard interladder
hoppings and we will also disregard the small 
interlayer hoppings t$_c$. Now we need
values for the onsite Coulomb interaction $U$ and the intersite Coulomb
interactions. 

Regarding $U$, it has been estimated in~\cite{yaresko00} to be $4.0$ eV and we
will use that value. The intersite Coulomb interaction can be estimated in the
following way. As shown 
in~\cite{thalmeier98} the Hamiltonian for the charge degrees of freedom
can be mapped to an antiferromagnetic Ising model in a transverse field. It is
well known~\cite{Pfeuty69} that in such a model there is a quantum critical
point at $2t_R=V_L$ and it has been argued that the material is close to this
point~\cite{mostovoy00}. 
Using $d_L=3.611$ {\AA} and $d_R=3.439$
{\AA} we can estimate $V_R$ by assuming that the
Coulomb interactions scale with $d^{-3}$ due to screening effects.

The results of our calculation are found in table~\ref{vvhop1}. The
estimate from experiment is $0.045\ldots 0.048$ eV~\cite{mila96,isobe96}, 
which agrees
with the results of our calculation. Using the LDA hopping
parameters~\cite{yaresko00} we find a value for 
J$_{eff}$ from the cluster calculation
which is too high. Using the parameters from the ab initio
calculation~\cite{suaud00}
we obtain a J$_{eff}$ which is slighly too low, but better than
LDA. Thus, the set of hopping parameters
obtained by the above treatment might be preferable for model calculations
on the ladders than using the LDA results, especially since the qualitative
features regarding $t_D$ and $t_L$ are conserved. 

\subsection {Hopping and superexchange in the distorted phase}
\label{sec:3.2}

We now repeat the above calculations for the distorted
phase: We obtain vanadium-oxygen, oxygen-sodium and vanadium-vanadium hoppings
by using the Slater-Koster method. 
 We then project out the vanadium-oxygen and the vanadium-sodium hoppings in
the single-particle Hamiltonian
to obtain the effective vanadium-vanadium hoppings. We calculate the
effective superexchange J$_{eff}$ for the V$_{1}$-ladders as we did for the
undistorted phase. However, we now 
have two inequivalent two-rung-clusters on this
ladder due to distortion. The effective superexchange between
electrons on neighbouring rungs is different for these two clusters. 
Therefore we
obtain an alternation of J$_{eff}$ along the ladder direction.
We did not calculate the
effective superexchange for the V$_{21}$-V$_{22}$-ladders,
since it is strongly modified by charge ordering.~\cite{yushankhai00} 
All results are found in table~\ref{resultsdist}.

\begin{table}
\caption {Effective vanadium-vanadium hoppings $t$, on-site-energy $\epsilon$
and superexchange J$_{eff}$ for the distorted phase given
in eV. Denomination of hoppings as given in figure~\ref{vanadiumladders}.
Superexchange J$_{eff}$ has been calculated with $U=4$ eV,
and $V_L=0.344$ as for the undistorted phase (see text).}
\label {resultsdist}
\begin {tabular} {lll}
\hline\noalign{\smallskip}
Hopping & Layer a & Layer b\\
\noalign{\smallskip}\hline\noalign{\smallskip}
\multicolumn {3}{c}{V$_{21}$-V$_{22}$-ladders}\\
t$_{R,CO}$ & 0.173 & 0.171\\
t$_{L,CO}$ & 0.052 & 0.057\\
t$_{D,CO,1}$ & 0.060 & 0.061\\
t$_{D,CO,2}$ & 0.084 & 0.086\\
t$_{cV_{21a}V_{21b}}$ & \multicolumn {2}{c} {0.015}\\
t$_{cV_{21a}V_{22b}}$ & \multicolumn {2}{c} {0.010}\\
t$_{cV_{22a}V_{21b}}$ & \multicolumn {2}{c} {0.005}\\
t$_{cV_{22a}V_{22b}}$ & \multicolumn {2}{c} {0.004}\\
$\epsilon_{V_{21}}$ & 1.265 & 1.250\\
$\epsilon_{V_{22}}$ & 1.421 & 1.428\\
\multicolumn {3}{c}{V$_{1}$-ladders}\\
t$_{R,SD}$ & 0.179 & 0.180\\ 
t$_{L,SD,1}$ & 0.062 & 0.068\\
t$_{L,SD,2}$ & 0.040 & 0.046\\
t$_{D,SD,1}$ & 0.070 & 0.070\\
t$_{D,SD,2}$ & 0.061 & 0.061\\
t$_{c,SD,1}$ &  \multicolumn {2}{c} {0.007}\\
t$_{c,SD,2}$ &  \multicolumn {2}{c} {0.006}\\
$\epsilon_{V_{1}}$ & 1.350 & 1.338\\
J$_{eff,1}$ & 0.066 & 0.072\\
J$_{eff,2}$ & 0.038 & 0.044\\
$\delta_J$ & 0.263 & 0.247\\
\multicolumn {3}{c}{Interladder}\\
t$_{IL1}$ & $-0.081$ & $-0.072$ \\
t$_{Il2}$ & $-0.130$ & $-0.128$\\
\noalign{\smallskip}\hline
\end {tabular}
\end {table}

For the V$_1$-ladders we find a large alternation of J$_{eff}$ along the
ladder with $\delta_J\approx 0.25$, 
larger by an order of magnitude than we had estimated from the change of
the vanadium-oxygen bond lengths. The dimerisation of $t_L$ along the ladder 
is $\delta_{t_L}\approx 0.2$ and therefore also much larger than what we
estimated from the variation of the bond length above. We also note that
$\delta_J$ and $\delta_{t_L}$ are of similar size, whereas we would expect a
factor of 2 difference from the standard superexchange estimate. 
The latter fact is,
however, easily explained: We estimate the effective superexchange between
electrons on neighbouring rungs of the same ladder by looking at the
singlet-triplet gap in such a two-rung cluster. This gap essentially describes
the kinetic energy gain associated with having the option of locating both
electrons on a single site, if they are in a singlet state, whereas they
cannot do so if they are in a triplet state. This means that all possible
two-particle states contribute to the size of the singlet-triplet gap,
including the states where both electrons are on a rung or sitting on the
diagonal. Therefore the standard estimate using 
$J=4t^2/U$ must fail and this is
also the reason for our definition of J$_{eff}$. This
explains why $\delta_J$ and $\delta_{t_L}$ are of similar size.

The reason for the unexpectedly strong alternation of both J$_{eff}$ and $t_L$
along the V$_{1}$-ladder of both layers is more complicated. To understand
this we take a closer look at how $t_L$ is actually derived from
the original vanadium-oxygen hopping terms and how these change under
distortion. Consider hopping along the leg from vanadium site V to V'
via the intermediate O$_1$ as shown in figure~\ref{oxygenpyramid}. The matrix
elements are given in table~\ref{hopleg}. We observe that due to the signs of
the hopping elements the effective hopping from V to V' via the O$_1$p$_y$
orbital counteracts the effective hopping via the O$_1$p$_x$ and the O$_1$p$_z$
orbital. Thus, in a simple approximation, which neglects all other
hoppings taking place in the system, the total effective hopping is

\begin{table}
\caption {Vanadium-oxygen hopping along the leg given in eV. 
Nomenclature as in
figure~\ref{oxygenpyramid}. The first two lines give the hopping for the
undistorted phase, the last four lines along the rungs of V$_{1a}$-ladder in
the distorted phase. 
}
\label {hopleg}
\begin {tabular} {llll}
\hline\noalign{\smallskip}
& t$_{p_xd_{xy}}$ & t$_{p_yd_{xy}}$ & t$_{p_zd_{xy}}$\\ 
\noalign{\smallskip}\hline\noalign{\smallskip}
\multicolumn{4}{l}{undistorted phase}\\
VO$_1$ & -0.8549 & 0.6303 & -0.2470\\
V'O$_1$ & 0.8549 & 0.6303 & 0.2470\\
\multicolumn{4}{l}{distorted phase: layer a}\\
VO$_1$ & -0.8486 & 0.6758 & -0.2534\\
V'O$_1$ & 0.8486 & 0.6758 & 0.2534\\
VO$_3$ & 0.8676 & 0.5860 & 0.2358\\
V''O$_3$ & -0.8676 & 0.5860 & -0.2358\\
\noalign{\smallskip}\hline
\end {tabular}
\end {table}

\begin {equation}
\label{tvv'}
t_{VV'} = -\frac{t_{p_xd_{xy}}^2}{\epsilon_{V}-\epsilon_{O}} + 
\frac{t_{p_yd_{xy}}^2}{\epsilon_{V}-\epsilon_{O}} -
\frac{t_{p_zd_{xy}}^2}{\epsilon_{V}-\epsilon_{O}} = -0.250\textrm{ eV}.
\end{equation}
\noindent
As we can also see from table~\ref{hopleg}, distortion affects the hoppings
via the different oxygen-p-orbitals in a significantly different way. 
For layer a the
dimerisation of these hoppings are approximately $\delta_x=0.01$,
$\delta_y=0.07$, $\delta_z=0.04$. But whenever $\left|t_{p_xd_{xy}}\right|$
decreases, $\left|t_{p_yd_{xy}}\right|$ increases and vice versa. Thus the
total dimerisation is enhanced: With the same crude approximation as in
equation (\ref{tvv'}) we obtain $t_{VV'}=-0.207$ eV and $t_{VV''}=-0.294$
eV. This gives a dimerisation of the effective vanadium-vanadium hopping of
$\delta_{t_{VV}}=0.17$. Thus we can explain the very large alternation of the
superexchange J$_{eff}$ which we obtained  for the V$_1$-ladders. Thus the
main result of the section is that this
alternation is almost completely due to the shift of the oxygen atoms on the
leg of the V$_1$ ladders in $a$ and $c$ direction, 
and is not caused by the small shift of the vanadium
sites.

We can also examine the behaviour of the hopping between layers. For the
spin-dimerized  
V$_1$ ladders the size of the effective hopping $t_{c,SD}$ changes
slightly, but there appears to be no significant alternation in c
direction. This is different for the effective hopping between charge ordered
ladders of neighbouring layers: Hopping is increased between V$_{21}$ sites
and decreased between V$_{22}$ sites. The size of the effective hopping
t$_{c,CO}$ remains very small however, causing a superexchange coupling of
less than 1 meV between completely localized electrons on the V$_{21}$
sites. We conclude that both in the undistorted and the distorted phase all
important spin interactions take place within a vanadium-oxygen layer.

\section {Summary}
\label{sec:4}

In this article we report a structure determination by synchrotron X-ray
diffraction of the low temperature phase of $\alpha$'-NaV$_2$O$_5$.
As in~\cite {ludecke99,deBoer00} two slightly different vanadium layers are
found to exist as well as two types of 
two-leg vanadium ladders within each layer. One of these ladders 
exhibits a zig-zag-type modulation and contains two inequivalent
vanadium sites V$_{21}$ and V$_{22}$. The other ladder contains only one
vanadium site V$_{1}$. 

Based on the structure determination we perform a
bond-valence calculation. Contrary to~\cite {smaalen00} we find \emph {only
two} different valences: the V$_{22}$ and the V$_{1}$ sites experience an
electron depletion towards a valence of V$^{4.5+\delta}$, 
while the charge on the V$_{21}$ sites is increased towards a valence of
V$^{4.5-3\delta}$. We 
compare our results with results of a $^{51}$V-NMR measurement~\cite
{ohama98} and find them to be in qualitatively good agreement: Due to
charge conservation an asymmetry both in the shift of the two new peaks from
the original V$^{4.5}$ peak  as well as an asymmetry in intensity are
found. 

We then use a Slater-Koster treatment to find the hopping along the bonds and
project out the V-O and V-Na processes
to obtain effective vanadium-vanadium hoppings.
The results compare well with LDA~\cite{yaresko00} as well as ab initio
calculations~\cite{suaud00} regarding relations between the different matrix
elements on the ladder. When we calculate the effective superexchange
between rungs for the undistorted phase using our matrix elements
we find a value of $0.047$ eV
which compares well with experimental results of $0.045\ldots 0.048$ eV. In
the distorted phase we obtain
a strong alternation of the superexchange with dimerisation $\delta_J\approx
0.25$ in the V$_1$ ladders along ladder direction. This dimerisation is 
nearly exclusively 
due to the shift of the oxygen atoms mediating the hopping along the legs
of the ladders in $a$ and $c$ direction; the shift of the vanadium atoms is much smaller. Even though
the Slater-Koster method is known to overestimate the effect of small
distortions, the results show clearly, that a significant alternation of the
superexchange is taking place on the V$_1$-ladders.   
No alternation in b-direction is found for the
V$_{21}$-V$_{22}$ ladders. No significant hopping or superexchange between
layers is found both for the undistorted and the distorted phase. 

This leaves
the question of the origin of the spin gap which has been observed in the
material. According to our results for the valence of the different vanadium
sites the picture of spin-singlet clusters proposed in~\cite{deBoer00} cannot
be correct, since it relies heavily on having V$^{5+}$ sites in the distorted
phase. However, in $^{23}$Na-NMR eight inequivalent sodium sites have been
detected~\cite{fagot99,ohama00}. In our structure determination we obtain only
six 
inequivalent sites. Four of these sites are located above and below the spin
dimerised V$_1$-ladders, two of them are located above and below the
charge-ordered V$_{21}$-V$_{22}$-ladders. To have eight inequivalent sodium
sites it would be sufficient to assume, that along the
V$_{21}$-V$_{22}$-ladders the shift of equivalent sodium
sites in c direction alternates along the ladder direction. This would also
immediately imply an alternation of the superexchange on the charge-ordered
ladders, since, as we have shown above, the hopping via the sodium site is
crucial to explain the diagonal hopping between two rungs of the same ladder.  
Alternatively, such an alternation of inequivalent sodium atoms can result
from an alternating shift of the rung oxygens of the V$_{21}$-V$_{22}$-ladders
in b direction. This would also cause a superexchange alternation along the 
V$_{21}$-V$_{22}$-ladders and one would also have eight inequivalent sodium
sites in the system. Such 
shifts would explain in a simple manner both the observation of $^{23}$Na-NMR
and the existence of a spin gap without disagreeing so far with the results
of all X-ray structure determinations. 
Since the superexchange alternation seems to be much
stronger on the V$_1$-ladders, this would imply a system with two
significantly different spin gaps. 

Another explanation for the observation of eight inequivalent sodium sites has
been given in~\cite{ohama00}. The authors propose that a charge ordering takes
place on all ladders. However, this implies having eight inequivalent sites
provided, one has a) an alternation of sodium sites along the ladder direction,
thus implying superexchange alternation as above for the case of the
V$_{21}$-V$_{22}$-ladders, b) two inequivalently charge ordered ladders. If
one had the same amount of charge ordering on both ladders one would also
expect the same sodium sites, since charge/valence and bonds are strongly
related. So, in such a case, four inequivalent vanadium sites would have to be
observed in $^{51}$V-NMR, which is not the case~\cite{ohama98}. 

In conclusion we have done a structure determination of the low temperature
phase of $\alpha'$NaV$_2$O$_5$. We confirm the results
of~\cite{ludecke99,deBoer00}, having charge ordered V$_{21}$-V$_{22}$-ladders
and disordered V$_1$-ladders. However, a bond valence calculation using our
results finds only two instead of three significantly different valences: The
valence of the  V$_1$ and the V$_{22}$ sites turns out to be nearly the
same. By making use of the Slater-Koster method we find a strong alternation
of the superexchange coupling between rungs of the V$_1$-ladders.

\begin{acknowledgement}
T. C. would like to thank Dr. H.-G. Krane for his help doing
synchrotron X-ray measurements. 
\end{acknowledgement}

\end {document}